\documentclass[11pt,a4paper]{article}
\usepackage{jcappub}

\usepackage{amsmath,amssymb}
\usepackage{graphicx}

\newcommand{\bk}{{\boldsymbol k}}

\newcommand{\bx}{{\boldsymbol x}}
\newcommand{\bB}{{\boldsymbol B}}

\newcommand{\beq}{\begin{equation}}
\newcommand{\eeq}{\end{equation}}
\newcommand{\ber}{\begin{eqnarray}}
\newcommand{\eer}{\end{eqnarray}}

\title{Reheating chiral dynamos with spin-0 and massive spin-1 torsions via chiral asymmetry}

\author[a]{Zhi-Fu Gao}
\author[a,b]{Biao-Peng Li}
\author[c,d]{L.C. Garcia de Andrade}

\affiliation[a]{Xinjiang Astronomical Observatory, Chinese Academy of Sciences, Urumqi, Xinjiang, 830011, China} %
\affiliation[b]{University of Chinese Academy of Sciences, Beijing, China} %
\affiliation[c]{Grupo de Gravitaç\~ao e Cosmologia-Departamento de
F\'{\i}sica Te\o'{o}rica - IF - UERJ - Rua S\~{a}o Francisco Xavier
524, Rio de Janeiro, RJ, Maracan\~{a}, CEP:20550} %
\affiliation[d]{Institute for Cosmology and Philosophy of Nature, Trg, Florjana, Croatia}%

\emailAdd{zhifugao@xao.ac.cn}
\emailAdd{libiaopeng@xao.ac.cn}
\emailAdd{luizandra795@gmail.com}

\abstract{Recently, Syderenko et al. (JCAP, 10: 018, 2016) investigated magnetogenesis and chiral asymmetry in the early hot universe. 
This study explores the impact of minimally coupling a constant torsion in their cosmological model, suggesting new chiral 
physics. Physically, this means that if torsion is right chiral, the difference between the number of right and left 
chiralities does not change. Moreover, the decay of chiral asymmetry depends on torsion chirality. We solve the chiral 
torsionful dynamo equation for magnetic field seeds. Magnetic helical fields are considered important for chiral fermion 
asymmetry. Even in $(3+1)$ dimensional spacetime, torsion is highly suppressed beyond inflation (Eur Phys J C 82: 291, 2022). 
However, torsion of $1\,\mathrm{MeV}$ appears in the early universe. Equations for correlated magnetic field coefficients are 
solved in terms of torsion. Weak magnetic fields of the order of $10^{-42}$ Gauss are boosted by powerful torsionful dynamo
 amplification, generating a much stronger magnetic field of the order of $10^{-9}$ Gauss in the present universe. A 
galactic magnetic field of $10^{-6}$ Gauss in the present universe, with torsion of $10^{-15}$ Gauss, leads us to a 
galactic dynamo seed of $10^{-9}$ Gauss. We also discuss reheating dynamo regeneration of decaying cosmic magnetic fields 
during the hadronization era. The relation between the reheating contribution to e-folds and the connection between CMF and 
temperature squared allows us to obtain dynamo amplification in terms of N-folds of inflation. The main innovation of this 
work is the exploration of constant torsion in a cosmological model, revealing new chiral physics. This study offers a new perspective 
on the origin and evolution of magnetic fields in the early universe.}


\arxivnumber{}

\begin{document}
\maketitle \flushbottom


\section{Introduction}

One of the major interests in modified gravity \cite{1} is that these theories, beyond Einstein's, show some advantages 
from both theoretical and experimental frameworks. These new theories often present features that Einstein's general 
relativity (GR) cannot explain satisfactorily \cite{2}.

In this paper, we aim to demonstrate that one of the most popular modifications of general relativity (GR), the Einstein-
Cartan (EC) gravity theory, exhibits such features. This theory was initially developed by Einstein and Cartan, and since 
then, it has received contributions from many researchers\cite{3,4,5,6,7,8,9,10,11,12,13,14,15,16,17}. Among the 
contributions to EC cosmology \cite{7,10,13,15,16,17}, Palle \cite{18} investigated torsion theories of gravity in 
cosmology, and obtained vorticity chiralities within the framework of EC and neutrino physics, using Goedel's stationary 
metrics. Palle showed that the spin density of light Majorana fermions acts as seeds for vorticity in the early universe, 
resulting in the universe's right-handed chirality. By utilizing rotational degrees of freedom, Palle \cite{18} applied a 
Goedel cosmological model in EC gravity, sourced by spin-density non-dynamical torsion. Chirality also leads to the spin 
flip of fermions when their chiralities are of the same species, either right or left-handed. It is crucial to investigate 
whether gauge fields and geometric anomalies affect this process. Previously, we also explored the chiral flipping of 
fermions in a constant torsion background.

Recently, Sydorenko et al.\cite{19} investigated the decay of chiral asymmetry in the early hot universe, discovering 
interesting results concerning the spin flip of fermions and the generation of non-zero magnetic helicity\cite{20,21}. 
Motivated by their work, and more recent findings on the influence of chiral torsion on massive photons in Proca geometry, 
we present new results on the effect of torsion on the decay of chiral asymmetry and dynamo instabilities in the early 
universe. These results highlight the role of chiral handedness in the EC universe.

This apparent new physics from torsion in EC gravity cosmological models strongly indicates the potential for torsion 
detection \cite{19} by comparing similar results in general relativistic cosmologies. Previous investigations of chiral 
dynamos in torsionful cyclic models\cite{23,24,25,26} have shown that torsion may also play an important role in mimicking 
the chiral chemical potential. Chiral dynamos were also studied within the framework of general relativistic cosmology (GRC) 
by Boyarsky et al.\cite{27}, who considered that the chemical chiral potential remains approximately constant at temperatures 
around 80 MeV.

More recently, Schober et al.\cite{28} addressed dynamo instability when the chiral chemical is non-uniform and inhomogeneous, 
performing numerical experiments. In our case, the difference between right and left-handed contexts would affect the chiral 
dynamo idea \cite{29}, where asymmetry can be converted between right and left-handed leptons in the early universe into 
helical magnetic fields. This idea proposes a possible magnetogenesis scenario. Helical magnetic fields can simplify the 
dynamo equations, and we also adopt the helical magnetogenesis hypothesis.

The cosmological model is sometimes non-minimally coupled to the electromagnetic field, but in this paper, we assume minimal 
coupling between torsion and the electromagnetic field is strictly observed. The decay of large-scale structure magnetic 
fields \cite{30} in the presence of torsion is obtained by considering the minimally coupled photon and torsion sectors. 
Consistent with Palle's result, we show from the chiral dynamo that in EC cosmology, chiral left-handed torsion seems to 
favor dynamo amplification.

Mavromatos \cite{31} demonstrated that a string-inspired Kalb-Ramond torsion field of 1 MeV is present in the 
early universe. This estimate is used here to obtain the torsion effects on magnetogenesis, showing a value of $10^{-45}$ 
Gauss(G) seed field to generate dynamo action from the observed astronomical value of $10^{-9}$ G in the present universe. 
Using a torsion field of 1 MeV, we see that the seed field must be very weak to align with astronomically observed 
late-universe magnetic fields. The early universe produces weak magnetic fields that are amplified by dynamo action, as 
discussed by Neronov and Semikoz \cite{29}. They also argued that chiral dynamo asymmetry converts fermion asymmetry into 
magnetic helicity in the early universe.

The remainder of this paper is organized as follows: Section\,\ref{sec2} investigates how chiral asymmetry in the presence of torsion 
modifies the sign of the difference between right and left-handed fermions. Section\,\ref{sec3} explores how the anomalous chiral 
magnetic effect (CME) is affected by torsion and determines magnetic correlation functions by solving their equations in a 
constant torsion background. Section\,\ref{sec4} is left for summary and discussion.

\section{Chiral asymmetry and chiral torsionful dynamo in Einstein-Cartan gravity}\label{sec2}
In this section, we will show how endowing spacetime with a torsion trace vector can significantly modify the chirality 
physics results in GR. We will briefly review the chiral asymmetry in the early hot universe and 
magnetic fields investigated by Siderenko et al.\cite{33}, because our work aims to embed their results and equations in a 
torsionful cosmology to explore new physics driven by torsion trace.

Let us start by considering a flat expanding universe filled with relativistic matter in the comoving coordinate system 
$({\eta},\bx)$, with conformal time coordinate ${\eta}$ and scale factor $a({\eta})$. The metric line element is 
given by
\begin{equation}
ds^{2}=a^{2}({\eta})(d{\eta}^{2}-d{\bx}^{2}).
\label{1}
\end{equation}
A divergence-free, statistically homogeneous, and isotropic cosmological magnetic field (CMF) has the following Fourier 
representation of the two-point correlation function:
\begin{equation}
<B_{i}(\bk),B^{*}_{j}({\bk}^{'})>= (2{\pi})^{3}{\delta}(\bk-\bk^{'})[P_{ij}S(k)+i\epsilon_{ijs}\hat{k}
^{s}A(k)].
\label{2}
\end{equation}
where $<B_{i}(\bk), B^{*}_{j}({\bk}')>$ describes the correlation between the magnetic field component $B_{i}$ 
at wave vector $\bk$ and the conjugate magnetic field component $B^{*}_{j}$ at the wave vector $\bk'$.
${i,j}=(1,2,3)$ represents the three spatial coordinates, and ${\hat{k}}^{s}=\hat{k}_{s}/\hat{k}$ represents the 
normalized unit vector component, $\delta(\hat{k} -\hat{k}')$ is the Dirac delta function, ensuring momentum conservation.
 $P_{ij}$ denotes the symmetric projector onto the plane orthogonal 
to $\bk$, defined as
\begin{equation}
P_{ij}={\delta}_{ij}- {\hat{k}}_{i}{\hat{k}}_{j} ,
\label{3}
\end{equation}
where ${\delta}_{ij}$ is the Kronecker delta function and ${\hat{k}}_{i}$, ${\hat{k}}_{j}$ 
are the normalized components of the wave vector. $S(k)$ is the scalar spectrum of the magnetic field, describing the 
energy density of the magnetic field. $i{\epsilon}_{ijs}{\hat{k}}^{s}A(k)$ represents the helicity component of the 
magnetic field, ${\epsilon}_{ijk}$ is the normalized totally skew-symmetric Levi-Civita symbol, which is used to describe 
the antisymmetric properties of the magnetic field. $A(k)$ is the spectrum describing the difference 
in power between left-handed and right-handed magnetic fields.

Now, let's introduce the helicity components $B_{\pm} (\bk)$ of the magnetic field in Fourier space, indexed by $i$,
\begin{equation}
B_{i}(\bk)=B_{+}(\bk)e_{i}^{+}({\bk})+B_{-}({\bk})e_{i}^{-}({\bk}),
\label{4}
\end{equation}
where the helicity components $B_{+}(\bk)$ and $B_{-}({\bk})$ correspond to the right-handed and left-handed 
circular polarization states, respectively, and $e_{i}^{+}({\bk})$ and $e_{i}^{-}({\bk})$ are complex basis 
vectors 
associated with these helicity components.
The complex basis $e_{i}^{\pm}$ is given by
\begin{equation}
e_{i}^{\pm}=\frac{1}{\sqrt{2}}(e^{1}_{i}{\pm}ie^{2}_{i}),
\label{5}
\end{equation}
formed from a right-handed and orthonormal basis $(e^{1}(\bk), e^{2}(\bk))$, with $e^{3}(\bk)$ 
proportional to the normalized wave vector ${\bk}$. 

The comoving magnetic field, $B$, is measured in a frame of reference that expands with the universe, while 
$B_{obs}$ is the magnetic field strength as observed at a specific point in time, without considering the effects of the 
universe's expansion. The factor $a$  represents the relative expansion of the universe. It changes over time, with $a$ 
being smaller in the past and larger in the present as the universe expands. The relationship between $B$ and 
 $B_{\rm obs}$ is $B = a^{2}B_{\rm obs}$, ensuring that the comoving magnetic field remains constant when we account for the 
expansion of the universe. In the case of a maximally helical magnetic field, the modulus (absolute value) of $A(k)$ is 
$|A(k)|= S(k)$, and the magnetic field is dominated by its left or right-handed part depending on the sign of $A(k)$.

The Abelian anomaly refers to the non-conservation of the axial current $J^\mu_5$ in the presence of electromagnetic fields 
in quantum field theory. It occurs due to quantum effects. The Abelian anomaly has important implications in various areas 
of physics, including the theory of the strong interaction (quantum chromodynamics), and plays a key role in the physics of 
the early universe and in the study of topological phases of matter. In this paper, we consider the effects of the Abelian anomaly in a torsionful constant spacetime and in the 
presence of spatially homogeneous chiral asymmetry. Thus, the evolution of the comoving magnetic field in 
conformal coordinates in a torsionless cosmic plasma with high conductivity ${\sigma}_{c}$ is
\begin{equation}
\frac{\partial \bB}{\partial \eta}=\sigma_{c}^{-1}{\nabla}^{2}{\bB}-\frac{{\alpha}{\Delta}_{\mu}}{{\pi}
{\sigma}_{c}}\nabla \times\bB, 
\label{6}
\end{equation}
where $\Delta \mu \equiv a \left({\mu}{L}-{\mu}{R}\right)$ is the spatially homogeneous difference between the (conformal) 
chemical potentials of the left-handed and right-handed charged leptons, $\sigma_c \equiv a \sigma \approx {\rm const.}$
\cite{34,35} characterizes the plasma conductivity, and $\alpha \approx 1/137$ is the fine structure constant. The first 
term on the left side of the equation represents the diffusion term, derived from Ohm's law in magnetohydrodynamics (MHD), 
and the last term is connected with the anomalous current in Maxwell's equations \cite{36,37,38,39,40}.

Recent papers have established a relationship between the chiral chemical potential and the 0-component of the 
torsion trace\cite{19,41,42,43}. Now, let us investigate the effects of introducing torsion in an otherwise flat spacetime by employing 
the minimal coupling
\begin{equation}
{D}_{0}= {\partial}_{\eta}-T_{0}.
\label{7}
\end{equation}
By introducing torsion, one can modify the dynamics of fields in spacetime, 
impacting the behavior of the magnetic field and its evolution. Let us rewrite the chiral dynamo equation, 
taking into account that the magnetic field is helical and obeys the equation
\begin{equation}
\nabla \times\bB={\lambda}{\bB}.
\label{8}
\end{equation}

Substituting this expression into the torsionless chiral dynamo equation implies
\begin{equation}
\frac{\partial \bB}{\partial \eta}={\sigma}_{c}^{-1}{\nabla}^{2}{\bB}-\frac{{\alpha}{\Delta}{\mu}}{{\pi}{\sigma}_{c}}
{\lambda}{\bB} .
\label{9}
\end{equation}
To further simplify this chiral dynamo equation, we propose a magnetic field ansatz such that ${\nabla}^{2}= -k$, 
whose substitution into expression\,(\ref{9}) yields
\begin{equation}
\frac{\partial B}{\partial \eta}=-\left(k{\sigma}_{c}^{-1}+\frac{{\alpha}{\Delta}{\mu}}
{{\pi}{\sigma}_{c}}{\lambda}\right )
{B}.
\label{10}
\end{equation}
This is a rather simple equation to be integrated. Now, by strategically coupling this equation to torsion through 
the derivative operator $D_{0}$, we get
\begin{equation}
\frac{{\partial}B}{{\partial}{\eta}}=-\left(-T_{0}+k{\sigma}_{c}^{-1}+\frac{{\alpha}{\Delta}{\mu}}{{\pi}{\sigma}_{c}}
{\lambda}\right)B.
\label{11}
\end{equation}
This is a first-order ordinary differential equation and can be integrated by separating the variables.

By rearranging  the variables and then integrating, we obtain
\begin{equation}
\int \frac{1}{B} dB = -\int \left(-T_{0} + k\sigma_{c}^{-1} + \frac{{\alpha}{\Delta}{\mu}}
{{\pi}\sigma_{c}}{\lambda}\right) d\eta .
\label{12}
\end{equation}
Performing the integration yields
\begin{equation}
 \mathrm{ln} B = \mathrm{constant} - \left(-T_{0} + k{\sigma}_{c}^{-1} + 
\frac{{\alpha}{\Delta}{\mu}}{{\pi}\sigma_{c}}{\lambda}\right) \eta .
\label{13}
\end{equation}
Exponentiating both sides,  we get a solution for $B$,
\begin{equation}
{B}({\eta})={B}_{0}\exp \left[(T_{0}-k{\sigma}_{c}^{-1}-\frac{{\alpha}{\Delta}{\mu}}{{\pi}
{\sigma}_{c}}{\lambda}){\eta}\right] ,
\label{14}
\end{equation}
where $B_{0}=B({\eta} \approx 0)$ is the initial magnetic field. 

Regarding the origin of cosmology magnetic fields, there are many theoretical hypotheses, one of which is the 
dynamo theory \cite{44,45,46,47,48,49,50,51,52,53,54}. The dynamo in general turbulent or a fast dynamo has 
to be very efficient at the ealy universe to amplify magnetic fields \cite{49,50,51,52,53,54,55,56}. The solution\,(\ref{14}) 
can be used to investigate dynamo mechanism with torsion for cosmology magnetic fields.

Notably, this solution produces some straightforward results. Near the singularity but with nonzero ${\eta}$, given the 
high conductivity of the early universe, we can simplify this expression to
\begin{equation}
{B}({\eta})={B}_{0}\exp\left[(T_{0}-\frac{{\alpha}{\Delta}{\mu}}{{\pi}{\sigma}_{c}}{\lambda}){\eta}\right] .
\label{15}
\end{equation}
To ensure that the magnetic field is amplified, the mathematical requirement is 
\begin{equation}
T_{0}-\frac{{\alpha}{\Delta}{\mu}}{{\pi}{\sigma}_{c}}{\lambda}> 0.
\label{16}
\end{equation}
Mathematically, this means right-handed or positive $T_{0}$ implies
\begin{equation}
{\Delta}{\mu}\ge{0}.
\label{17}
\end{equation}
From Equation\,(\ref{17}), it is shown that the magnetic field ${B}({\eta})$ evolves exponentially with conformal time $\eta$, 
with the growth rate determined by the torsion parameter $T_{0}$, wave vector $k$, conductivity ${\sigma}_{c}$, and the 
chiral chemical potential ${\Delta}{\mu}$. In the above expression, even with constant torsion trace, the chirality of torsion induces a dominance of one chirality over the 
other. Specifically, if the torsion is right-handed or positive, left-handed particles in the Einstein-Cartan universe 
dominate over right-handed particles. It is important to emphasize that Equation\,(\ref{15}) shows that the onset of the 
dynamo mechanism is favored when the chiral torsion is right-handed and the chiral chemical potential variation is positive. 

\section{Magnetic field correlation via chiral torsion}\label{sec3}
In this section, we shall address the role of torsion chirality in the decay of chiral asymmetry and solve Sydorenko 
et al's equations \cite{18} for magnetic field correlation coefficients to obtain this correlation in terms of torsion chirality. 
Using Equations\,(\ref{2}) and\,(\ref{6}), one can obtain the following system of equations for
the spectra $S(k,\eta)$ and $A(k,\eta)$ (see Ref.\cite{27}]):
\begin{equation} 
\frac{\partial S}{\partial \eta} =\left[T_{0}S - \frac{2k^{2}}{{\sigma}_{c}}(S - S_{{\rm eq}})\right] + \frac{2
\alpha k}{\pi \sigma_c} \Delta \mu A ,
\label{18} 
\end{equation} 
and 
\begin{equation} 
\frac{\partial A}{\partial \eta} = \left(T_{0}A - \frac{2k^{2}}{{\sigma}_{c}}A + \frac{2{\alpha}k{\Delta}{\mu}}{{\pi}{\sigma}_{c}}\right)S .
\label{19} 
\end{equation}
These equations are derived in the torsionless case by Sydorenko et al.\,\cite{19} by using the chiral dynamo equation and 
correlation field. These equations can be further simplified if one considers that not only is the conductivity very 
high in the early universe (almost a superconductor). In Equation (\ref{20}), we included a term representing the thermal 
equilibrium distribution
\begin{equation} 
S_{\mathrm{eq}} = \frac{k}{\exp[\frac{k}{aT}] - 1} ,
\label{20} 
\end{equation}
which serves to ensure that the spectral energy distribution $S$ relaxes to its equilibrium value  $S_{\rm eq}$ instead of 
going to zero. However, this mechanism does not apply in the long-wavelength domain $k \lesssim 1/\eta$, which is not causally connected in the expanding hot universe. 
To simplify matters, we will omit the term $S_{\mathrm{eq}}$ and since in the long wavelength 
limit $k^{2} \approx 0$, Equations\,(\ref{20}) and\,(\ref{21}) reduce to
\begin{equation} 
\frac{\partial S}{\partial \eta} = T_{0}S + \frac{2{\alpha}k{\Delta}{\mu}}{{\pi}
{\sigma}_{c}}A .
\label{21} 
\end{equation}

Now it is easy to solve this system of correlation functions by multiplying Equation\,(\ref{24}) by $A$,
\begin{equation} 
A \frac{\partial S}{\partial \eta} = T_{0}SA + \frac{2{\alpha}k{\Delta}{\mu}}{{\pi}{\sigma}_{c}}A^{2}, 
\label{22} 
\end{equation}
and multiplying Equation\,(\ref{21}) by $S$ yields
\begin{equation} 
S \frac{\partial A}{\partial \eta} = T_{0}AS - \frac{2{\alpha}k{\Delta}{\mu}}{{\pi}{\sigma}_{c}}S^{2} .
\label{23} 
\end{equation}
Therefore, adding these equations yields
\begin{equation} 
\frac{\partial(AS)}{\partial\eta} = 2T_{0}(AS) + \frac{2{\alpha}k{\Delta}{\mu}}{{\pi}{\sigma}_{c}}(A^{2} + S^{2}) .
\label{24} 
\end{equation}
Assuming that $A$ and $S$ are small quantities, one obtains the new result that does not appear in the absence of torsion 
\begin{equation} 
\frac{\partial(AS)}{\partial\eta} \approx 2T_{0}(AS) .
\label{25} 
\end{equation}
This is an extremely simple differential equation that can easily be solved to yield
\begin{equation} 
AS = (AS)_{0} \exp(2T_{0}{\eta}) .
\label{26} 
\end{equation}
However, this expression does not completely determine the values of $A$ and $S$. In this particular case, they can be 
equal, resulting in
\begin{equation} 
S^{2} = S^{2}_{0} \exp(2T_{0}{\eta}) .
\label{27} 
\end{equation}
The final expressions describe the exponential growth of the product $AS$ and the individual spectra $S$ in the presence 
of torsion $T_0$.

With the constraint $S=A$, the two-point correlation function in terms of torsion becomes
\begin{equation} 
<B_{i}(\bk),{B^{*}}{j}({\bk}')> = (2{\pi})^{3}{\delta}(\bk - \bk')\left[P_{ij} + 
i{\epsilon}_{ijs}{\hat{k}}^{s}\right] A(k).
\label{28} 
\end{equation}

Finally, the expression for the two-point correlation function of magnetic fields in terms of chiral torsion is
\begin{equation} 
<B_{i}(\bk),{B^{*}}{j}({\bk}')>= (2{\pi})^{3}{\delta}(\bk - \bk')\left[P_{ij} + 
i{\epsilon}_{ijs}{\bk}^{s}\right]\exp[T_{0}k] .
\label{29} 
\end{equation}
The projection operator $P_{ij}$ and the pseudoscalar term $i{\epsilon}_{ijs}{k}^{s}$ correctly capture 
the isotropic and helical components of the magnetic field correlation.
From the dynamical equation above, we obtain a wave-like equation for the function $A(k)$ as follows
\begin{equation} 
\frac{\partial^2 A}{\partial \eta^2} = \left[T_{0}-\frac{2k^{2}}{{\sigma}_{c}}\frac{\partial A}{\partial \eta}+\frac{2{\alpha}k}{{\pi}{\sigma}_{c}}\frac{\partial\Delta\mu}{\partial\eta} \right] S .
\label{30} 
\end{equation}
This wave-like equation for $A$ is easily solved if we assume that
\begin{equation} 
\frac{\partial\Delta\mu}{\partial\eta} \approx 0 . 
\label{31} 
\end{equation}

Substituting Equation\,(\ref{31}) into Equation\,(\ref{30}) yields
\begin{equation} 
\frac{\partial^2 A}{\partial \eta^2} - \left(T_{0}-\frac{2k^{2}}{{\sigma}_{c}}\right) 
\frac{\partial A}{\partial \eta} = 0 .
\label{32} 
\end{equation}
As a result, we get the following solution
\begin{equation} 
A(k) = 2A_{0} \left(T_{0} + \frac{2k{2}}{\sigma_{c}}\right)^{-1} 
\exp\left[2\left(T_{0} - \frac{2k^{2}}{{\sigma}_{c}}\right){\eta}\right] .
\label{33} 
\end{equation}
Therefore, this solution can also be damped by a strong left-handed chiral torsion. Let us now examine the case of 
the decay of the chiral chemical potential or chiral asymmetry ${\Delta}{\mu}$. In a flat space, we have
\begin{equation} 
\frac{d{\Delta}{\mu}}{d{\eta}} = 
-\frac{c_{\Delta}}{{\pi}^{2}} \int \left[-T(k,{\eta})A + \frac{\partial A}{\partial\eta}\right] k\, dk .
\label{34} 
\end{equation}

By taking torsion as constant and substituting the value of $A$ in its simplest function of torsion, one obtains
\begin{equation} 
\frac{d{\Delta}{\mu}}{d{\eta}}= -\frac{c_{\Delta}}{{\pi}^{2}} A_{0} (1 - T_{0}k^{2}) \exp(T_{0}{\eta}) .
\label{35} 
\end{equation}
Here, $c_\Delta$ is a numerical constant of order unity (it would be equal to $3/4$ in
pure quantum electrodynamics) that reflects the particle content of the primordial
plasma \cite{19}.

Integrating this expression with respect to ${\eta}$ yields the following result
\begin{equation} 
{\Delta}{\mu} = -\frac{c_{\Delta}}{{\pi}^{2}}A_{0}(1 - T_{0}k^{2})\int_{0}^{\eta}\exp(T_{0}{\eta})d{\eta}.
\label{36} 
\end{equation}
\begin{figure}[h]
    \centering
    \includegraphics[width=0.7\linewidth]{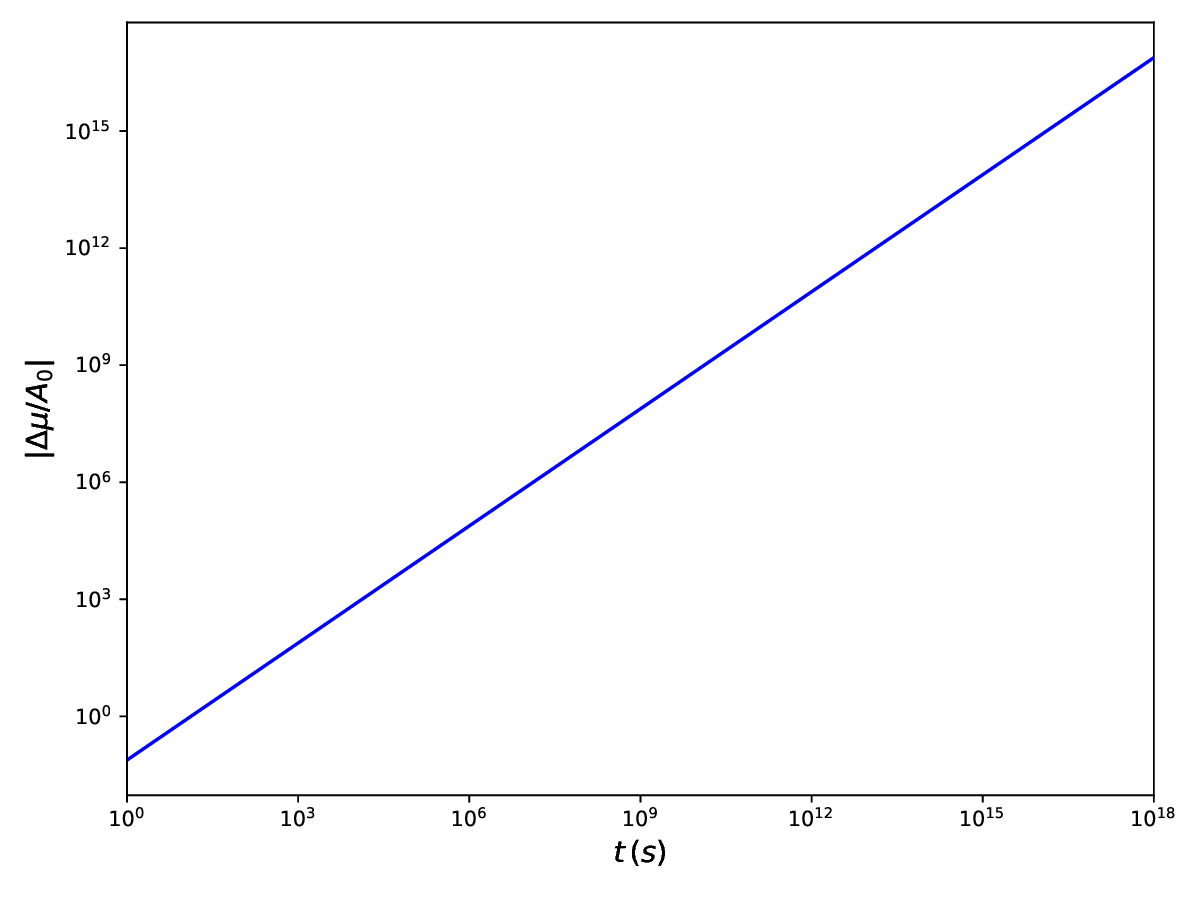}
    \caption{The relation between chiral chemical potential $|\Delta \mu /A_0|$ and cosmic time $t$. It is derived from the Equation\,(\ref{37}). The numerical values are plotted across the range of cosmic times, from $1$ s to $10^{18}$ s.}
    \label{fig1}
\end{figure}
If we take a left-handed chiral torsion, the difference between left and right-handed fermions is enhanced by the presence 
of torsion. Thus, left-handed torsion implies that the chiral asymmetry decays. For long wavelengths, the term 
$T_{0}k^{2}$ is too small and can be neglected in Equation (\ref{36}), leading to a simplified expression for the difference 
between left and right chiral fermions ${\Delta}{\mu}$ as
\begin{equation} 
{\Delta}{\mu} \approx  -\frac{c_{\Delta}}{{\pi}^{2}}A_{0} \frac{\exp(T_{0}t)-1}{T_{0}},
\label{37} 
\end{equation}
where we have substituted the conformal time by the cosmic time $t$ because spacetime is nearly flat at the Earth's surface. 
For torsion at the surface of the Earth, we may use the value obtained by Bergmann and de Sabbata \cite{50} of $10^{-24}$ 
s$^{-1}$. Although this is a significant small value of $T_{0}$, if we take the cosmic time $t=10^{18}$\,s, its effect  
on chirality flip is significant, $T_{0}t \sim 10^{12}$.  The evolution of the chiral chemical potential ${\Delta}{\mu}$ 
depends on the coefficients $c_{\Delta}/\pi^{2}$, $A_0$, $T_{0}$ and time $t$. Since the value of $A_{0}$ is hard to be 
evaluated, we may normalize the chiral chemical potential as ${\Delta}{\mu}/A_{0}$ and make a plot of  ${\Delta}{\mu}/A_{0}$ 
as a function of $t$ (Figure\,\ref{fig1}).

\section{Summary and Discussion}\label{sec4}
The study explores the problem of chiral magnetic effects in torsionful spacetime. According to Mavromatos and Mavromatos
\cite{31}, such effects are not possible in string-inspired theory. By assuming minimal coupling between torsion and 
magnetic fields, we obtain chiral asymmetry from torsion by solving the dynamo equation and the correlation function for 
magnetic fields at two distinct points in spacetime within torsionful spacetime cosmology. The Einstein-Cartan gravity in its classical format was 
used, and torsion was introduced through the covariant derivative with torsion. Our findings show that the introduction of 
torsion changes the sign of chiral asymmetry and plays a fundamental role in the decay of the chiral chemical potential.

Our approach differs from that of Shaposhnikov et al.\cite{33}, who use totally skew-symmetric axial torsion, whereas we assume the 
torsion trace only contribution. The advantage of their Palatini-like formulation is that it links chiral asymmetry to 
chiral anomalies, both axial and gravitational. Simplification of our equations is achieved by assuming a long wavelength, 
which allows the wave vector $\bk$ to be neglected when squared. This assumption enables us to solve the equations for 
the two-point functions easily. 

Also, using Equation\,(\ref{15}), we can estimate the amplification factor of $B$ as $\xi={B}({\eta})/{B}_{0}={B}({\eta})/{B}_{\rm seed}$. For example, if 
taking $B$ as the $10^{-9}$ G as the observed magnetic field at present universe and ${B}_{\rm seed}\sim 
10^{-24}$\,G as the seed magnetic field at the early universe\,\cite{48}, $\xi$ is estimated as the order of $10^{15}$. 
However, we cannot directly provide the value of the "seed" magnetic field in the early universe through observations. 
This magnetic field value often relies on theoretical models. Even if an exact value for ${B}_{\rm seed}$ is given, 
it is challenging to determine the observational value of $T_{0}$ due to several reasons: (1) many parameters in the 
equation (such as conductivity $\sigma_{c}$, helicity $\lambda$, dynamo amplification timescale $\eta$) are unknown; 
(2) the equation itself is derived through certain assumptions and simplifications. 

Future research should provide a more detailed account of dynamo instability from chiral asymmetry in cosmological models 
that can accommodate torsion. One intriguing argument to pursue further is the fact that while most particles have 
antiparticles, left-handed neutrinos do not. This leads us to consider whether our Einstein-Cartan left-handed universe 
could be sourced by left-handed neutrinos, which are fermions that couple with torsion. The main issue here is that we use 
a frame where only the time component of torsion is non-vanishing, making it difficult to associate neutrino helicity with 
axial torsion.

An important point to address is that torsion is highly suppressed not only in braneworld models but also in the 3+1 
spacetime of Einstein-Cartan-Brans-Dicke inflation \cite{22}. Another interesting perspective for future research is to 
investigate whether the results presented here favor Einstein-Cartan cosmology, as we have recently shown in the case of 
the anti-de Sitter universe\cite{51}.
\acknowledgments

We would like to thank D Sokolov for helpful discussions on the subject of Riemannian dynamos. Luiz thanks his  
wife Ana Paula Teixeira Araujo for her constant support. Financial support from UERJ is gratefully ackownoledged.
This work was performed under the auspices of Major Science and Technology Program of Xinjiang Uygur Autonomous Region 
through No.2022A03013-1.

\end{document}